\documentclass[%
 reprint,
 amsmath,amssymb,
 aps,
 prb,
]{revtex4-1}

\usepackage{graphicx}

\begin{document}
\title{Scattering from spin-polarized charged impurities in graphene}

\author{Ville Vierimaa}
 \email{ville.v.vierimaa@aalto.fi}
\author{Zheyong Fan}
\author{Ari Harju}
\affiliation{COMP Centre of Excellence, Department of Applied Physics, Aalto University, Helsinki, Finland}

\date{\today}

\begin{abstract}

We study the spin relaxation of charge carriers in graphene in the presence of spin-polarized charged impurities by calculating the time evolution of initially polarized state. The spin relaxation time shows completely different energy behaviour for short-ranged and long-ranged spin scatterers and can be used to identify the dominant source of spin scattering. Our results agree well with recent experimental findings and indicate that their spin relaxation is likely caused by long-ranged scatterers. 
\end{abstract}

\maketitle

%\section{Introduction}
Graphene is one of the most promising materials for spintronics. It has a small intrinsic spin-orbit coupling and early theoretical work predicted it to have spin relaxation time of up to 1 $\mu$s \cite{Ertler2009, Han2014}. This value is however yet to be reached, as typical experiments report spin relaxation times between hundred picoseconds and a few nanoseconds \cite{Popinciuc2009, Han2012, Swartz2013, Yang2011, Maassen2012}, recently reaching 12 ns \cite{Drogeler2016}. Recent theoretical studies have uncovered the origin of the shorter-than-expected relaxation times to some extent \cite{vanTuan2014}, but more study on the underlying mechanism is still needed.

Most of the theories on spin relaxation have included the spin coupling either in the form of a Rashba term in the Hamiltonian or by adding magnetic defects in the system. The Rashba Hamiltonian is a coupling between spin and the momentum, induced by an electric field perpendicular to the graphene plane \cite{Rashba1960, Dedkov2008, Marchenko2012, Weeks2011}. Magnetic defects, on the other hand, are impurities, in which the spin coupling is caused by the finite magnetic moment of the defects. The magnetic defects come in multiple varieties and their size can range from point defects, such as adatoms or vacancies \cite{Yazyev2007, Hong2012}, to large-scale defects such as edge states \cite{Nakada1996, Fujita1996}. It has been demonstrated that the magnetic defects can be described by adding a spin-dependent potential term to the tight-binding Hamiltonian \cite{Thomsen2015, Ervasti2015, Fujita1996}. 

In this work, we focus on magnetic defects and study charged impurities, in which the potential is spin-dependent. Charged impurities, also called electron-hole puddles, have been studied quite extensively for charge transport in graphene \cite{Martin2008, Adam2009, Zhang2009}. However, their effect on the spin relaxation has not been studied much and the focus has been more on the Rashba-type coupling \cite{vanTuan2016}. In contrast to the conventional magnetic defects, the electron-hole puddles span multiple sites in the graphene lattice and can model a variable range of defects. This is an ideal model for studying the difference between short- and long-ranged scatterers, allowing us to show that spin relaxation in experiments is most likely caused by long-ranged scatterers.

%\section{Model}
We model the electron-hole puddles as Gaussian-shaped potential fluctuations. A system of pristine graphene is used as a starting point for the calculations, described by a tight-binding Hamiltonian
\begin{equation}
H_0=-t_0\sum_{\langle i,j\rangle} |i\rangle\langle j|,
\end{equation}
where $\langle i,j\rangle$ denotes the set of nearest neighbours in the system and $t_0=2.7$ eV is the hopping between the neighbours. The potential for each defect is given by $U(r)=U_0^s e^{-r^2/2r_0^2}$, with $U_0^s$ being the potential strength for spin $s$ and $r$ the distance from the potential center. This simple defect model can be used for both short- and long-ranged scatterers by varying the defect width $r_0$. In the limit of $r_0$ going to zero a simple on-site potential is recovered. The potential strength for each defect is determined by choosing an average potential $U_{\text{ave}}$ randomly from $[-U_{\text{max}},U_{\text{max}}]$ and adding (subtracting) a spin splitting $\Delta$ to arrive at the potential for spin up (down). Therefore, for spin up we have $U_0^\uparrow=U_{\text{ave}} + \Delta$ and for spin down $U_0^\downarrow=U_{\text{ave}} - \Delta$. Writing the potentials in this way lets us express the defect Hamiltonian for a single defect centered at $\vec{r}_d$ as \cite{Ervasti2015}
\begin{equation}
H_d = \sum_i |i\rangle\langle i|\otimes \left(U_{\text{ave}} I + \Delta \sigma_d \right)e^{-|\vec{r}_d-\vec{r}_i|^2/2r_0^2},
\end{equation}
where $I$ is an identity matrix in the spin basis and $\sigma_d$ is a rotated Pauli $z$-matrix. For a general defect, we can write
\begin{equation}
\sigma_d=\begin{bmatrix}
\cos \theta & e^{i\phi}\sin \theta \\ e^{-i\phi}\sin \theta & -\cos \theta
\end{bmatrix},
\end{equation}
where the angles $\theta$ and $\phi$ refer to the orientation of the defect on the Bloch sphere. 

When we consider multiple defects in the simulation, they are assumed not to interact with each other. This means that they do not alter each other's parameters and their Hamiltonians can be simply added together, allowing us to write the total Hamiltonian as
\begin{equation}
H=H_0+\sum_d H_d.
\end{equation}
To quantify the number of defects in the system, we define the defect density $\rho$ as the ratio between the number of defect centres and the number of atoms in the system. 

%%%%%%%%%%%%%%%%%%%%%%%%%%%%%%%%%%%%%%%%%%%%%%%%%%%%%%%%%%%%%%%%%%%%%%%%%%%%%%%
%\section{Methods}

The velocity autocorrelation function $C_{vv}$ is an important quantity in the Kubo-Greenwood formalism because the conductivity is given by its integral \cite{Kubo1957}. However, we are more interested in $C_{vv}$ itself because it also contains information about the time scale of charge relaxation. The function is defined as an energy-projected average from 
\begin{equation}
C_{vv}(E,t)=\frac
{\text{Tr}\left\lbrace  V(t)V \delta(E-H)\right\rbrace}
{\text{Tr}\left\lbrace  \delta(E-H)\right\rbrace},
\label{vac_definition}
\end{equation}
where $V= i[H,X]/\hbar$ is the velocity operator and $V(t)=U^\dagger(t)VU(t)$ is its representation in the Heisenberg picture, $U(t)$ being the time evolution operator. The projection to energy $E$ is done by the delta function $\delta(E-H)$. To study the spin relaxation, we use a similar method. A natural way of looking into spin is to calculate the spin polarization, which can be calculated\cite{vanTuan2014} by replacing the velocity operators in Eq.(\ref{vac_definition}) by the Pauli $z$-matrix $s_z(t)=U^\dagger(t)s_zU(t)$,
\begin{equation}
S(E,t)=\frac{\text{Tr}\left\lbrace s_z(t) \delta(E-H) \right\rbrace}{\text{Tr}\left\lbrace\delta(E-H)\right\rbrace}.
\label{S_definition}
\end{equation}
In principle, the spin polarization could be calculated as a vector for all components, but the relaxation time can be determined from the component corresponding to the initial polarization, which we have defined to be the z-direction.

The traces in Eqs. (\ref{vac_definition}) and (\ref{S_definition}) scale poorly with the system size. Also, the delta-functions and time-evolution cannot be expressed in a closed form, which means that the two equations cannot be used directly. Instead, we apply a series of approximations commonly used in linear-scaling Kubo-Greenwood conductivity calculations\cite{Fan2014, Settnes2016, Ferreira2015, Garcia2015}. The first and most important approximation is to replace the trace with a sum over random-phase states \cite{Weisse2006}. This makes the calculation linear-scaling and allows us to reach large enough system size to eliminate the finite-size effects caused by rather large defects we have. The other approximations done are mostly technical because the time evolution and delta function cannot be calculated analytically. Instead, they are evaluated numerically using a Chebyshev expansion \cite{Weisse2006, TalEzer1984, Fehske2009}. For the delta-function, the expansion is done up to 3000 Chebyshev moments. This gives a half-maximum width of 2 meV for the delta-function and is accurate enough for our purposes. In the time-evolution, the accuracy of the expansion depends on the time-step used and a fixed number of moments cannot be chosen. Instead, the expansion is terminated when the magnitude of expansion coefficients drops below $10^{-15}$.

We study the charge and spin relaxation by calculating the velocity autocorrelation function and the spin polarization as a function of time. Starting from a random initial state, both quantities decay towards zero and their time-evolution behaviours can be used to extract relaxation times. We assume the decay to be exponential and obtain the relaxation time $\tau$ by fitting $A e^{-t/\tau}$ to the calculated data. From here on, we will denote the charge relaxation time as $\tau_v$ and the spin relaxation time as $\tau_s$. The relaxation needs not be exponential \cite{Thomsen2015}, but the exponential fit should give a good approximation on the relaxation time regardless. The velocity autocorrelation also has a dampening oscillatory part in it, but the simple least-squares fitting captures the decay relatively well. For the spin polarization, there are only small deviations from the exponential behaviour in the systems we have considered.

%\section{Results}

\begin{figure}
\includegraphics[width=\linewidth]{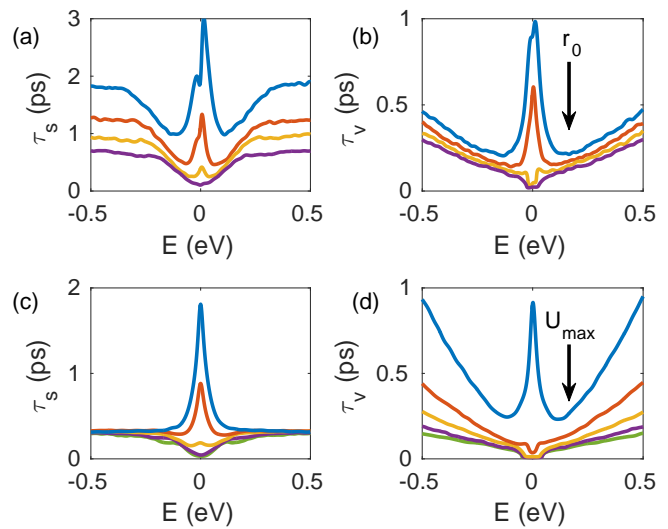}
\caption{(Color online) Spin (a,c) and charge (b,d) relaxation times as a function of energy. In (a,b) $r_0$ is varied from 1.1 nm to 1.6 nm while keeping $U_{\text{max}}$ at 0.7 eV and in (c,d) $U_{\text{max}}$ goes from 0.27 eV to 0.81 eV while $r_0$ stays at 2.1 nm. Both parameters are increased with even spacing and the arrows indicate the direction of increasing $r_0$ and $U_{\text{max}}$. For all cases, $\rho$ is kept at 50 ppm and $\Delta$ at 0.1 eV.}
\label{range_transition}
\end{figure}

\begin{figure}
\includegraphics[width=\linewidth]{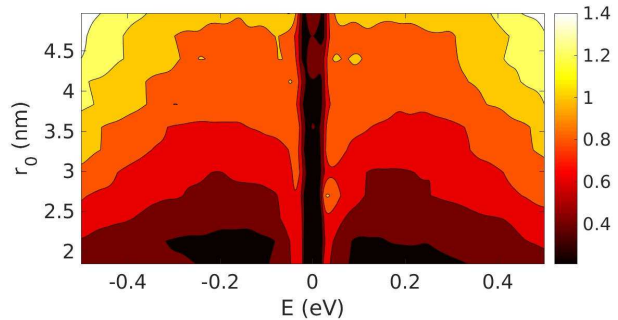}
\caption{(Color online) The ratio $\tau_v/\tau_s$ as a function of defect width and energy with $U_\text{max}$=0.7 eV, $\Delta=0.1$ eV and $\rho=10$ ppm.}
\label{relaxation_scaling}
\end{figure}

Our model displays a transition when the defect range is varied. In the short-range limit both $\tau_s$ and $\tau_v$ are peaked at the Dirac point. As the defect range is increased, the peak changes to a minimum abruptly, especially for $\tau_v$, as seen in Fig. \ref{range_transition}. This transition seems to be related to the crossover between ballistic and diffusive transport. In the small defect limit, the velocity autocorrelation has strong oscillations at $E=0$ even at longer correlation times. Conductivity, which is proportional to the integral of the velocity autocorrelation, shows no signs of saturation within the simulation time of 5 ps and strongly hints to ballistic transport. With larger defects, $C_{vv}$ decays almost completely during the simulation time and its integral will saturate, indicating diffusive behaviour. Similar behaviour is also observed when the defect potential $U_{\text{max}}$ is varied.

A curious observation from Figs. \ref{range_transition}(b) and \ref{range_transition}(d) is the constant plateau in $\tau_v$ that can be seen around the Dirac point in the case of the largest defects. It is present in all long-ranged test cases we have considered and it is caused by the relatively large value we have used for $\Delta$. The plateau gets narrower when $\Delta$ is decreased and it is most likely absent when $\Delta$ reaches an experimentally relevant value. It will however affect some of our results near the Dirac point. 

Spin relaxation also shows a similar transition as charge relaxation, even though its behaviour is slightly different. The peak changes to a minimum during the transition much more smoothly and the energy dependence is also different. While the charge relaxation time increases linearly away from the Dirac point, spin relaxation will eventually tend to a constant value. This behaviour is mostly explained by the magnitude of the defect potential compared to the charge carrier's energy. As the charge carrier energy is increased, the carriers can move through the defects more easily and eventually the defects do not affect their movement at all. At that point, the relaxation rate of the spin is completely determined by the effective concentration of the defects and does not change with carrier energy.

In experiments, the size of the charged impurities in graphene samples can be tens of nanometers \cite{Zhang2009, Xue2011}. This size range is far beyond the transition region we considered earlier and we can expect both relaxation times to be similar in the sense that both of them have a minimum at the Dirac point. Because they are similar, it makes sense to consider the ratio $\tau_v/\tau_s$ to see if they scale the same way. This ratio is shown in Fig. \ref{relaxation_scaling}. When the scatterer range is short, $\tau_v/\tau_s$ is quite small and charge relaxes much faster than spin. As the scatterer range is increased, the ratio gets larger and larger and $\tau_s$ becomes shorter than $\tau_v$ quite fast. This behaviour is universal across the energy, except for a narrow region around the Dirac point, in which $\tau_v/\tau_s$ grows quite slowly. This different region corresponds to the plateau in $\tau_v$  and will most likely get even narrower when $\Delta$ is decreased. In general, the spin relaxation seems to be fastest at higher energies and with longer-ranged scatterers.

\begin{figure*}
\includegraphics[width=\linewidth]{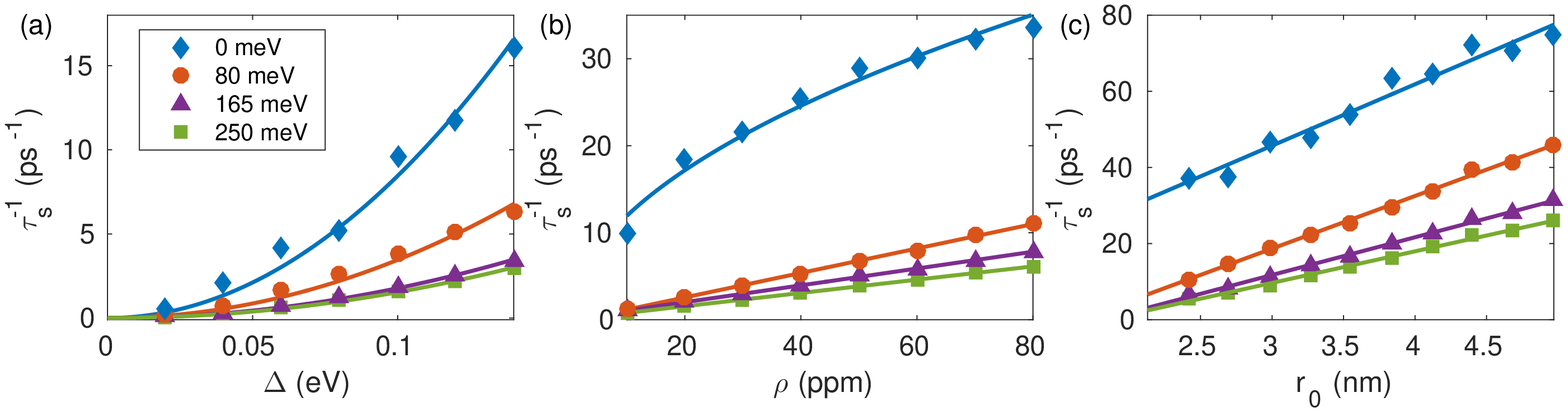}
\caption{(Color online) Scaling of $\tau_s$ as a function of (a) spin splitting, (b) defect density and (c) scatterer range. The solid lines show fitted curves, which are quadratic in (a) and linear in (b) and (c), except for the $E=0$ case in (b) where a power law fit has been used.}
\label{scaling}
\end{figure*}

To compare our results with experiments, we need to calculate the scaling of the spin relaxation time with respect to the defect parameters, because our $\Delta$ is much larger than one would expect the spin splitting to be in experiments. The other parameters are already in the experimentally viable range, but it is convenient to have the scaling relation for them also. As seen already in Fig. \ref{range_transition}, $U_\text{max}$ has only a small effect on $\tau_s$ when $r_0$ is relatively large and we have left it out of the scaling analysis. The scaling of $\tau_s^{-1]}$ with respect to the remaining three defect parameters, $\Delta$, $\rho$ and $r_0$, is shown in Fig. \ref{scaling}.

The conventional mechanisms for spin relaxation \cite{Ochoa2012, Dyakonov1972} predict a quadratic dependence between $\tau_s^{-1}$ and $\Delta$ \cite{Boross2013}. The fits to $\tau_s^{-1} \sim \Delta^2$ shown in Fig. \ref{S_definition}(a) indicate that this is also the case with our model. There seems to be some noise in the data at lower energies, but overall the agreement is quite nice. Based on this, a $\tau_s^{-1} \sim \Delta^2$ scaling can be expected to work quite well especially at higher energies. Near the Dirac point there can be some deviations and the results from low energies should be treated carefully.

For $\rho$ and $r_0$ the scaling is mostly linear as seen in Figs. \ref{scaling}(b) and \ref{scaling}(c). The only exception is found at the Dirac point for $\rho$ where the scaling is closer to $\rho^{-1/2}$. For higher energies the linear fit represents the data well. We also note that $\tau_s^{-1}$ should go to zero along with both $\rho$ and $r_0$ because there is no spin scattering in the absence of any defects. Extrapolating the data to $\rho=0$ agrees well with this condition and the behaviour seems to be linear all the way. For $r_0$ the behaviour is slightly more complicated because the range transition gives non-trivial scaling for shorter-ranged defects.

\begin{figure}
\includegraphics[width=0.85\linewidth]{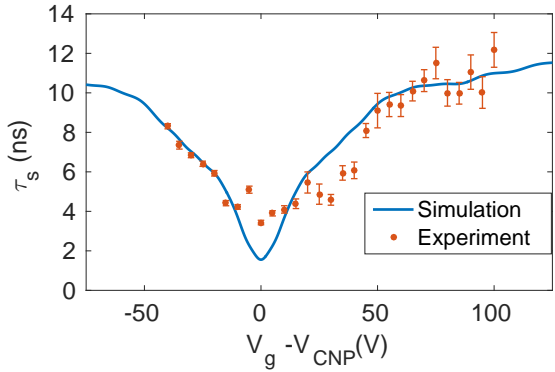}
\caption{(Color online) Calculated spin relaxation time compared to experiments by Dr\"{o}geler \textit{et al.} \cite{Drogeler2016}. The defect parameters for the simulations are $r_0$ = 4.1 nm, $U_\text{max}$ = 0.7 eV and $\rho$ = 10 ppm. The results have been calculated with $\Delta$ = 0.1 eV, which has been scaled down to 410 $\mu$eV.}
\label{tau_s_comparison}
\end{figure}

In Fig. \ref{tau_s_comparison} we show a comparison between our calculated spin relaxation time and the experiments by Dr\"{o}geler \textit{et al}\cite{Drogeler2016}. The defect parameters (see Fig. \ref{tau_s_comparison} caption) for our calculation have been chosen to get a good match in the energy dependence and $\Delta$ has been scaled down to match the magnitude of $\tau_s$. We have also transformed our energy to carrier density by integrating the DOS and carrier density to gate voltage through $n=\alpha (V_\text{g}-V_{\text{CNP}})$, where $\alpha=4.8\times 10^{10}$ V$^{-1}$cm$^{-2}$ is the capacitive coupling to the backgate in the experiment \cite{Drogeler2016}. The charge neutrality point $V_{\text{CNP}}$ is taken to be the voltage at which $\tau_s$ gets its smallest value, $V_\text{g}=30$ V.

The match between the two spin relaxation times is really good, especially at larger energies. The differences at small energies are expected because the Dirac point behaves slightly differently. Based on the scaling we have applied to $\Delta$, we can also give an approximation on the spin coupling in the experiments. If we assume that $\Delta$ would be the only parameter that needs to be scaled down, we can use the quadratic scaling to arrive at $\Delta=410~\mu$eV, which is a rather large value. However, the spin coupling is usually assumed to be uniform across the system while we have a non-uniform coupling in our calculations. To give a more fair comparison to uniform coupling, we average $\Delta$ over the system to get an effective uniform coupling $\Delta_{eff}=17$ $\mu$eV. This is comparable to usual values of intrinsic spin-orbit coupling\cite{Konschuh2010, Abdelouahed2010, Boettger2007}.

%\section{Conclusions}
In conclusion, we have shown that spin-polarized charged impurities provide a reasonable model for studying spin relaxation in graphene. The energy dependence of the spin relaxation time given by the model is strikingly similar to the recent experimental measurements \cite{Han2012, Yang2011, Drogeler2016} and suggests that the spin relaxation in these experiments is caused by long-ranged scatterers. Magnitude of spin relaxation time is comparable to the experiments, as long as the spin splitting is scaled down to relevant range.

\begin{acknowledgments}
We thank M. Dr\"{o}geler, C. Stampfer and B. Beschoten for providing their experimental data and helpful comments. This work was supported by the Academy of Finland through its Centres of Excellence Programme (2015-2017) under project number 284621. We acknowledge the computational resources provided by Aalto Science-IT project and Finland’s IT Center for Science (CSC).
\end{acknowledgments}


\bibliography{Spin_scatterers_in_graphene}

\begin{thebibliography}{38}%
\makeatletter
\providecommand \@ifxundefined [1]{%
 \@ifx{#1\undefined}
}%
\providecommand \@ifnum [1]{%
 \ifnum #1\expandafter \@firstoftwo
 \else \expandafter \@secondoftwo
 \fi
}%
\providecommand \@ifx [1]{%
 \ifx #1\expandafter \@firstoftwo
 \else \expandafter \@secondoftwo
 \fi
}%
\providecommand \natexlab [1]{#1}%
\providecommand \enquote  [1]{``#1''}%
\providecommand \bibnamefont  [1]{#1}%
\providecommand \bibfnamefont [1]{#1}%
\providecommand \citenamefont [1]{#1}%
\providecommand \href@noop [0]{\@secondoftwo}%
\providecommand \href [0]{\begingroup \@sanitize@url \@href}%
\providecommand \@href[1]{\@@startlink{#1}\@@href}%
\providecommand \@@href[1]{\endgroup#1\@@endlink}%
\providecommand \@sanitize@url [0]{\catcode `\\12\catcode `\$12\catcode
  `\&12\catcode `\#12\catcode `\^12\catcode `\_12\catcode `\%12\relax}%
\providecommand \@@startlink[1]{}%
\providecommand \@@endlink[0]{}%
\providecommand \url  [0]{\begingroup\@sanitize@url \@url }%
\providecommand \@url [1]{\endgroup\@href {#1}{\urlprefix }}%
\providecommand \urlprefix  [0]{URL }%
\providecommand \Eprint [0]{\href }%
\providecommand \doibase [0]{http://dx.doi.org/}%
\providecommand \selectlanguage [0]{\@gobble}%
\providecommand \bibinfo  [0]{\@secondoftwo}%
\providecommand \bibfield  [0]{\@secondoftwo}%
\providecommand \translation [1]{[#1]}%
\providecommand \BibitemOpen [0]{}%
\providecommand \bibitemStop [0]{}%
\providecommand \bibitemNoStop [0]{.\EOS\space}%
\providecommand \EOS [0]{\spacefactor3000\relax}%
\providecommand \BibitemShut  [1]{\csname bibitem#1\endcsname}%
\let\auto@bib@innerbib\@empty
%</preamble>
\bibitem [{\citenamefont {Ertler}\ \emph {et~al.}(2009)\citenamefont {Ertler},
  \citenamefont {Konschuh}, \citenamefont {Gmitra},\ and\ \citenamefont
  {Fabian}}]{Ertler2009}%
  \BibitemOpen
  \bibfield  {author} {\bibinfo {author} {\bibfnamefont {C.}~\bibnamefont
  {Ertler}}, \bibinfo {author} {\bibfnamefont {S.}~\bibnamefont {Konschuh}},
  \bibinfo {author} {\bibfnamefont {M.}~\bibnamefont {Gmitra}}, \ and\ \bibinfo
  {author} {\bibfnamefont {J.}~\bibnamefont {Fabian}},\ }\href@noop {}
  {\bibfield  {journal} {\bibinfo  {journal} {Phys. Rev. B}\ }\textbf {\bibinfo
  {volume} {80}},\ \bibinfo {pages} {041405} (\bibinfo {year}
  {2009})}\BibitemShut {NoStop}%
\bibitem [{\citenamefont {Han}\ \emph {et~al.}(2014)\citenamefont {Han},
  \citenamefont {Kawakami}, \citenamefont {Gmitra},\ and\ \citenamefont
  {Fabian}}]{Han2014}%
  \BibitemOpen
  \bibfield  {author} {\bibinfo {author} {\bibfnamefont {W.}~\bibnamefont
  {Han}}, \bibinfo {author} {\bibfnamefont {R.~K.}\ \bibnamefont {Kawakami}},
  \bibinfo {author} {\bibfnamefont {M.}~\bibnamefont {Gmitra}}, \ and\ \bibinfo
  {author} {\bibfnamefont {J.}~\bibnamefont {Fabian}},\ }\href@noop {}
  {\bibfield  {journal} {\bibinfo  {journal} {Nat. Nano}\ }\textbf {\bibinfo
  {volume} {9}},\ \bibinfo {pages} {794} (\bibinfo {year} {2014})}\BibitemShut
  {NoStop}%
\bibitem [{\citenamefont {Popinciuc}\ \emph {et~al.}(2009)\citenamefont
  {Popinciuc}, \citenamefont {J\'ozsa}, \citenamefont {Zomer}, \citenamefont
  {Tombros}, \citenamefont {Veligura}, \citenamefont {Jonkman},\ and\
  \citenamefont {van Wees}}]{Popinciuc2009}%
  \BibitemOpen
  \bibfield  {author} {\bibinfo {author} {\bibfnamefont {M.}~\bibnamefont
  {Popinciuc}}, \bibinfo {author} {\bibfnamefont {C.}~\bibnamefont {J\'ozsa}},
  \bibinfo {author} {\bibfnamefont {P.~J.}\ \bibnamefont {Zomer}}, \bibinfo
  {author} {\bibfnamefont {N.}~\bibnamefont {Tombros}}, \bibinfo {author}
  {\bibfnamefont {A.}~\bibnamefont {Veligura}}, \bibinfo {author}
  {\bibfnamefont {H.~T.}\ \bibnamefont {Jonkman}}, \ and\ \bibinfo {author}
  {\bibfnamefont {B.~J.}\ \bibnamefont {van Wees}},\ }\href@noop {} {\bibfield
  {journal} {\bibinfo  {journal} {Phys. Rev. B}\ }\textbf {\bibinfo {volume}
  {80}},\ \bibinfo {pages} {214427} (\bibinfo {year} {2009})}\BibitemShut
  {NoStop}%
\bibitem [{\citenamefont {Han}\ \emph {et~al.}(2012)\citenamefont {Han},
  \citenamefont {Chen}, \citenamefont {Wang}, \citenamefont {McCreary},
  \citenamefont {Wen}, \citenamefont {Swartz}, \citenamefont {Shi},\ and\
  \citenamefont {Kawakami}}]{Han2012}%
  \BibitemOpen
  \bibfield  {author} {\bibinfo {author} {\bibfnamefont {W.}~\bibnamefont
  {Han}}, \bibinfo {author} {\bibfnamefont {J.-R.}\ \bibnamefont {Chen}},
  \bibinfo {author} {\bibfnamefont {D.}~\bibnamefont {Wang}}, \bibinfo {author}
  {\bibfnamefont {K.~M.}\ \bibnamefont {McCreary}}, \bibinfo {author}
  {\bibfnamefont {H.}~\bibnamefont {Wen}}, \bibinfo {author} {\bibfnamefont
  {A.~G.}\ \bibnamefont {Swartz}}, \bibinfo {author} {\bibfnamefont
  {J.}~\bibnamefont {Shi}}, \ and\ \bibinfo {author} {\bibfnamefont {R.~K.}\
  \bibnamefont {Kawakami}},\ }\href@noop {} {\bibfield  {journal} {\bibinfo
  {journal} {Nano Lett.}\ }\textbf {\bibinfo {volume} {12}},\ \bibinfo {pages}
  {3443} (\bibinfo {year} {2012})}\BibitemShut {NoStop}%
\bibitem [{\citenamefont {Swartz}\ \emph {et~al.}(2013)\citenamefont {Swartz},
  \citenamefont {Chen}, \citenamefont {McCreary}, \citenamefont {Odenthal},
  \citenamefont {Han},\ and\ \citenamefont {Kawakami}}]{Swartz2013}%
  \BibitemOpen
  \bibfield  {author} {\bibinfo {author} {\bibfnamefont {A.~G.}\ \bibnamefont
  {Swartz}}, \bibinfo {author} {\bibfnamefont {J.-R.}\ \bibnamefont {Chen}},
  \bibinfo {author} {\bibfnamefont {K.~M.}\ \bibnamefont {McCreary}}, \bibinfo
  {author} {\bibfnamefont {P.~M.}\ \bibnamefont {Odenthal}}, \bibinfo {author}
  {\bibfnamefont {W.}~\bibnamefont {Han}}, \ and\ \bibinfo {author}
  {\bibfnamefont {R.~K.}\ \bibnamefont {Kawakami}},\ }\href@noop {} {\bibfield
  {journal} {\bibinfo  {journal} {Phys. Rev. B}\ }\textbf {\bibinfo {volume}
  {87}},\ \bibinfo {pages} {075455} (\bibinfo {year} {2013})}\BibitemShut
  {NoStop}%
\bibitem [{\citenamefont {Yang}\ \emph {et~al.}(2011)\citenamefont {Yang},
  \citenamefont {Balakrishnan}, \citenamefont {Volmer}, \citenamefont {Avsar},
  \citenamefont {Jaiswal}, \citenamefont {Samm}, \citenamefont {Ali},
  \citenamefont {Pachoud}, \citenamefont {Zeng}, \citenamefont {Popinciuc},
  \citenamefont {G\"untherodt}, \citenamefont {Beschoten},\ and\ \citenamefont
  {\"Ozyilmaz}}]{Yang2011}%
  \BibitemOpen
  \bibfield  {author} {\bibinfo {author} {\bibfnamefont {T.-Y.}\ \bibnamefont
  {Yang}}, \bibinfo {author} {\bibfnamefont {J.}~\bibnamefont {Balakrishnan}},
  \bibinfo {author} {\bibfnamefont {F.}~\bibnamefont {Volmer}}, \bibinfo
  {author} {\bibfnamefont {A.}~\bibnamefont {Avsar}}, \bibinfo {author}
  {\bibfnamefont {M.}~\bibnamefont {Jaiswal}}, \bibinfo {author} {\bibfnamefont
  {J.}~\bibnamefont {Samm}}, \bibinfo {author} {\bibfnamefont {S.~R.}\
  \bibnamefont {Ali}}, \bibinfo {author} {\bibfnamefont {A.}~\bibnamefont
  {Pachoud}}, \bibinfo {author} {\bibfnamefont {M.}~\bibnamefont {Zeng}},
  \bibinfo {author} {\bibfnamefont {M.}~\bibnamefont {Popinciuc}}, \bibinfo
  {author} {\bibfnamefont {G.}~\bibnamefont {G\"untherodt}}, \bibinfo {author}
  {\bibfnamefont {B.}~\bibnamefont {Beschoten}}, \ and\ \bibinfo {author}
  {\bibfnamefont {B.}~\bibnamefont {\"Ozyilmaz}},\ }\href {\doibase
  10.1103/PhysRevLett.107.047206} {\bibfield  {journal} {\bibinfo  {journal}
  {Phys. Rev. Lett.}\ }\textbf {\bibinfo {volume} {107}},\ \bibinfo {pages}
  {047206} (\bibinfo {year} {2011})}\BibitemShut {NoStop}%
\bibitem [{\citenamefont {Maassen}\ \emph {et~al.}(2012)\citenamefont
  {Maassen}, \citenamefont {van~den Berg}, \citenamefont {Ijbema},
  \citenamefont {Fromm}, \citenamefont {Seyller}, \citenamefont {Yakimova},\
  and\ \citenamefont {van Wees}}]{Maassen2012}%
  \BibitemOpen
  \bibfield  {author} {\bibinfo {author} {\bibfnamefont {T.}~\bibnamefont
  {Maassen}}, \bibinfo {author} {\bibfnamefont {J.~J.}\ \bibnamefont {van~den
  Berg}}, \bibinfo {author} {\bibfnamefont {N.}~\bibnamefont {Ijbema}},
  \bibinfo {author} {\bibfnamefont {F.}~\bibnamefont {Fromm}}, \bibinfo
  {author} {\bibfnamefont {T.}~\bibnamefont {Seyller}}, \bibinfo {author}
  {\bibfnamefont {R.}~\bibnamefont {Yakimova}}, \ and\ \bibinfo {author}
  {\bibfnamefont {B.~J.}\ \bibnamefont {van Wees}},\ }\href@noop {} {\bibfield
  {journal} {\bibinfo  {journal} {Nano Lett.}\ }\textbf {\bibinfo {volume}
  {12}},\ \bibinfo {pages} {1498} (\bibinfo {year} {2012})}\BibitemShut
  {NoStop}%
\bibitem [{\citenamefont {Dr\"{o}geler}\ \emph {et~al.}(2016)\citenamefont
  {Dr\"{o}geler}, \citenamefont {Franzen}, \citenamefont {Volmer},
  \citenamefont {Pohlmann}, \citenamefont {Banszerus}, \citenamefont {Wolter},
  \citenamefont {Watanabe}, \citenamefont {Taniguchi}, \citenamefont
  {Stampfer},\ and\ \citenamefont {Beschoten}}]{Drogeler2016}%
  \BibitemOpen
  \bibfield  {author} {\bibinfo {author} {\bibfnamefont {M.}~\bibnamefont
  {Dr\"{o}geler}}, \bibinfo {author} {\bibfnamefont {C.}~\bibnamefont
  {Franzen}}, \bibinfo {author} {\bibfnamefont {F.}~\bibnamefont {Volmer}},
  \bibinfo {author} {\bibfnamefont {T.}~\bibnamefont {Pohlmann}}, \bibinfo
  {author} {\bibfnamefont {L.}~\bibnamefont {Banszerus}}, \bibinfo {author}
  {\bibfnamefont {M.}~\bibnamefont {Wolter}}, \bibinfo {author} {\bibfnamefont
  {K.}~\bibnamefont {Watanabe}}, \bibinfo {author} {\bibfnamefont
  {T.}~\bibnamefont {Taniguchi}}, \bibinfo {author} {\bibfnamefont
  {C.}~\bibnamefont {Stampfer}}, \ and\ \bibinfo {author} {\bibfnamefont
  {B.}~\bibnamefont {Beschoten}},\ }\href@noop {} {\bibfield  {journal}
  {\bibinfo  {journal} {Nano Lett.}\ }\textbf {\bibinfo {volume} {16}},\
  \bibinfo {pages} {3533} (\bibinfo {year} {2016})}\BibitemShut {NoStop}%
\bibitem [{\citenamefont {Van~Tuan}\ \emph {et~al.}(2014)\citenamefont
  {Van~Tuan}, \citenamefont {Ortmann}, \citenamefont {Soriano}, \citenamefont
  {Valenzuela},\ and\ \citenamefont {Roche}}]{vanTuan2014}%
  \BibitemOpen
  \bibfield  {author} {\bibinfo {author} {\bibfnamefont {D.}~\bibnamefont
  {Van~Tuan}}, \bibinfo {author} {\bibfnamefont {F.}~\bibnamefont {Ortmann}},
  \bibinfo {author} {\bibfnamefont {D.}~\bibnamefont {Soriano}}, \bibinfo
  {author} {\bibfnamefont {S.~O.}\ \bibnamefont {Valenzuela}}, \ and\ \bibinfo
  {author} {\bibfnamefont {S.}~\bibnamefont {Roche}},\ }\href
  {http://dx.doi.org/10.1038/nphys3083} {\bibfield  {journal} {\bibinfo
  {journal} {Nat Phys}\ }\textbf {\bibinfo {volume} {10}},\ \bibinfo {pages}
  {857} (\bibinfo {year} {2014})}\BibitemShut {NoStop}%
\bibitem [{\citenamefont {Rashba}(1960)}]{Rashba1960}%
  \BibitemOpen
  \bibfield  {author} {\bibinfo {author} {\bibfnamefont {E.~I.}\ \bibnamefont
  {Rashba}},\ }\href@noop {} {\bibfield  {journal} {\bibinfo  {journal} {Sov.
  Phys. Solid. State}\ }\textbf {\bibinfo {volume} {2}},\ \bibinfo {pages}
  {1109} (\bibinfo {year} {1960})}\BibitemShut {NoStop}%
\bibitem [{\citenamefont {Dedkov}\ \emph {et~al.}(2008)\citenamefont {Dedkov},
  \citenamefont {Fonin}, \citenamefont {R\"udiger},\ and\ \citenamefont
  {Laubschat}}]{Dedkov2008}%
  \BibitemOpen
  \bibfield  {author} {\bibinfo {author} {\bibfnamefont {Y.~S.}\ \bibnamefont
  {Dedkov}}, \bibinfo {author} {\bibfnamefont {M.}~\bibnamefont {Fonin}},
  \bibinfo {author} {\bibfnamefont {U.}~\bibnamefont {R\"udiger}}, \ and\
  \bibinfo {author} {\bibfnamefont {C.}~\bibnamefont {Laubschat}},\ }\href@noop
  {} {\bibfield  {journal} {\bibinfo  {journal} {Phys. Rev. Lett.}\ }\textbf
  {\bibinfo {volume} {100}},\ \bibinfo {pages} {107602} (\bibinfo {year}
  {2008})}\BibitemShut {NoStop}%
\bibitem [{\citenamefont {Marchenko}\ \emph {et~al.}(2012)\citenamefont
  {Marchenko}, \citenamefont {Varykhalov}, \citenamefont {Scholz},
  \citenamefont {Bihlmayer}, \citenamefont {Rashba}, \citenamefont {Rybkin},
  \citenamefont {Shikin},\ and\ \citenamefont {Rader}}]{Marchenko2012}%
  \BibitemOpen
  \bibfield  {author} {\bibinfo {author} {\bibfnamefont {D.}~\bibnamefont
  {Marchenko}}, \bibinfo {author} {\bibfnamefont {A.}~\bibnamefont
  {Varykhalov}}, \bibinfo {author} {\bibfnamefont {M.~R.}\ \bibnamefont
  {Scholz}}, \bibinfo {author} {\bibfnamefont {G.}~\bibnamefont {Bihlmayer}},
  \bibinfo {author} {\bibfnamefont {E.~I.}\ \bibnamefont {Rashba}}, \bibinfo
  {author} {\bibfnamefont {A.}~\bibnamefont {Rybkin}}, \bibinfo {author}
  {\bibfnamefont {A.~M.}\ \bibnamefont {Shikin}}, \ and\ \bibinfo {author}
  {\bibfnamefont {O.}~\bibnamefont {Rader}},\ }\href@noop {} {\bibfield
  {journal} {\bibinfo  {journal} {Nat. Commun}\ }\textbf {\bibinfo {volume}
  {3}},\ \bibinfo {pages} {1232} (\bibinfo {year} {2012})}\BibitemShut
  {NoStop}%
\bibitem [{\citenamefont {Weeks}\ \emph {et~al.}(2011)\citenamefont {Weeks},
  \citenamefont {Hu}, \citenamefont {Alicea}, \citenamefont {Franz},\ and\
  \citenamefont {Wu}}]{Weeks2011}%
  \BibitemOpen
  \bibfield  {author} {\bibinfo {author} {\bibfnamefont {C.}~\bibnamefont
  {Weeks}}, \bibinfo {author} {\bibfnamefont {J.}~\bibnamefont {Hu}}, \bibinfo
  {author} {\bibfnamefont {J.}~\bibnamefont {Alicea}}, \bibinfo {author}
  {\bibfnamefont {M.}~\bibnamefont {Franz}}, \ and\ \bibinfo {author}
  {\bibfnamefont {R.}~\bibnamefont {Wu}},\ }\href@noop {} {\bibfield  {journal}
  {\bibinfo  {journal} {Phys. Rev. X}\ }\textbf {\bibinfo {volume} {1}},\
  \bibinfo {pages} {021001} (\bibinfo {year} {2011})}\BibitemShut {NoStop}%
\bibitem [{\citenamefont {Yazyev}\ and\ \citenamefont
  {Helm}(2007)}]{Yazyev2007}%
  \BibitemOpen
  \bibfield  {author} {\bibinfo {author} {\bibfnamefont {O.~V.}\ \bibnamefont
  {Yazyev}}\ and\ \bibinfo {author} {\bibfnamefont {L.}~\bibnamefont {Helm}},\
  }\href {\doibase 10.1103/PhysRevB.75.125408} {\bibfield  {journal} {\bibinfo
  {journal} {Phys. Rev. B}\ }\textbf {\bibinfo {volume} {75}},\ \bibinfo
  {pages} {125408} (\bibinfo {year} {2007})}\BibitemShut {NoStop}%
\bibitem [{\citenamefont {Hong}\ \emph {et~al.}(2012)\citenamefont {Hong},
  \citenamefont {Zou}, \citenamefont {Wang}, \citenamefont {Cheng},\ and\
  \citenamefont {Zhu}}]{Hong2012}%
  \BibitemOpen
  \bibfield  {author} {\bibinfo {author} {\bibfnamefont {X.}~\bibnamefont
  {Hong}}, \bibinfo {author} {\bibfnamefont {K.}~\bibnamefont {Zou}}, \bibinfo
  {author} {\bibfnamefont {B.}~\bibnamefont {Wang}}, \bibinfo {author}
  {\bibfnamefont {S.-H.}\ \bibnamefont {Cheng}}, \ and\ \bibinfo {author}
  {\bibfnamefont {J.}~\bibnamefont {Zhu}},\ }\href {\doibase
  10.1103/PhysRevLett.108.226602} {\bibfield  {journal} {\bibinfo  {journal}
  {Phys. Rev. Lett.}\ }\textbf {\bibinfo {volume} {108}},\ \bibinfo {pages}
  {226602} (\bibinfo {year} {2012})}\BibitemShut {NoStop}%
\bibitem [{\citenamefont {Nakada}\ \emph {et~al.}(1996)\citenamefont {Nakada},
  \citenamefont {Fujita}, \citenamefont {Dresselhaus},\ and\ \citenamefont
  {Dresselhaus}}]{Nakada1996}%
  \BibitemOpen
  \bibfield  {author} {\bibinfo {author} {\bibfnamefont {K.}~\bibnamefont
  {Nakada}}, \bibinfo {author} {\bibfnamefont {M.}~\bibnamefont {Fujita}},
  \bibinfo {author} {\bibfnamefont {G.}~\bibnamefont {Dresselhaus}}, \ and\
  \bibinfo {author} {\bibfnamefont {M.~S.}\ \bibnamefont {Dresselhaus}},\
  }\href@noop {} {\bibfield  {journal} {\bibinfo  {journal} {Phys. Rev. B}\
  }\textbf {\bibinfo {volume} {54}},\ \bibinfo {pages} {17954} (\bibinfo {year}
  {1996})}\BibitemShut {NoStop}%
\bibitem [{\citenamefont {Fujita}\ \emph {et~al.}(1996)\citenamefont {Fujita},
  \citenamefont {Wakabayashi}, \citenamefont {Nakada},\ and\ \citenamefont
  {Kusakabe}}]{Fujita1996}%
  \BibitemOpen
  \bibfield  {author} {\bibinfo {author} {\bibfnamefont {M.}~\bibnamefont
  {Fujita}}, \bibinfo {author} {\bibfnamefont {K.}~\bibnamefont {Wakabayashi}},
  \bibinfo {author} {\bibfnamefont {K.}~\bibnamefont {Nakada}}, \ and\ \bibinfo
  {author} {\bibfnamefont {K.}~\bibnamefont {Kusakabe}},\ }\href@noop {}
  {\bibfield  {journal} {\bibinfo  {journal} {J. Phys. Soc. Jpn.}\ }\textbf
  {\bibinfo {volume} {65}},\ \bibinfo {pages} {1920} (\bibinfo {year}
  {1996})}\BibitemShut {NoStop}%
\bibitem [{\citenamefont {Thomsen}\ \emph {et~al.}(2015)\citenamefont
  {Thomsen}, \citenamefont {Ervasti}, \citenamefont {Harju},\ and\
  \citenamefont {Pedersen}}]{Thomsen2015}%
  \BibitemOpen
  \bibfield  {author} {\bibinfo {author} {\bibfnamefont {M.~R.}\ \bibnamefont
  {Thomsen}}, \bibinfo {author} {\bibfnamefont {M.~M.}\ \bibnamefont
  {Ervasti}}, \bibinfo {author} {\bibfnamefont {A.}~\bibnamefont {Harju}}, \
  and\ \bibinfo {author} {\bibfnamefont {T.~G.}\ \bibnamefont {Pedersen}},\
  }\href {\doibase 10.1103/PhysRevB.92.195408} {\bibfield  {journal} {\bibinfo
  {journal} {Phys. Rev. B}\ }\textbf {\bibinfo {volume} {92}},\ \bibinfo
  {pages} {195408} (\bibinfo {year} {2015})}\BibitemShut {NoStop}%
\bibitem [{\citenamefont {Ervasti}\ \emph {et~al.}(2015)\citenamefont
  {Ervasti}, \citenamefont {Fan}, \citenamefont {Uppstu}, \citenamefont
  {Krasheninnikov},\ and\ \citenamefont {Harju}}]{Ervasti2015}%
  \BibitemOpen
  \bibfield  {author} {\bibinfo {author} {\bibfnamefont {M.~M.}\ \bibnamefont
  {Ervasti}}, \bibinfo {author} {\bibfnamefont {Z.}~\bibnamefont {Fan}},
  \bibinfo {author} {\bibfnamefont {A.}~\bibnamefont {Uppstu}}, \bibinfo
  {author} {\bibfnamefont {A.~V.}\ \bibnamefont {Krasheninnikov}}, \ and\
  \bibinfo {author} {\bibfnamefont {A.}~\bibnamefont {Harju}},\ }\href
  {\doibase 10.1103/PhysRevB.92.235412} {\bibfield  {journal} {\bibinfo
  {journal} {Phys. Rev. B}\ }\textbf {\bibinfo {volume} {92}},\ \bibinfo
  {pages} {235412} (\bibinfo {year} {2015})}\BibitemShut {NoStop}%
\bibitem [{\citenamefont {Martin}\ \emph {et~al.}(2008)\citenamefont {Martin},
  \citenamefont {Akerman}, \citenamefont {Ulbricht}, \citenamefont {Lohmann},
  \citenamefont {Smet}, \citenamefont {von Klitzing},\ and\ \citenamefont
  {Yacoby}}]{Martin2008}%
  \BibitemOpen
  \bibfield  {author} {\bibinfo {author} {\bibfnamefont {J.}~\bibnamefont
  {Martin}}, \bibinfo {author} {\bibfnamefont {N.}~\bibnamefont {Akerman}},
  \bibinfo {author} {\bibfnamefont {G.}~\bibnamefont {Ulbricht}}, \bibinfo
  {author} {\bibfnamefont {T.}~\bibnamefont {Lohmann}}, \bibinfo {author}
  {\bibfnamefont {J.~H.}\ \bibnamefont {Smet}}, \bibinfo {author}
  {\bibfnamefont {K.}~\bibnamefont {von Klitzing}}, \ and\ \bibinfo {author}
  {\bibfnamefont {A.}~\bibnamefont {Yacoby}},\ }\href {\doibase
  10.1038/nphys781} {\bibfield  {journal} {\bibinfo  {journal} {Nat. Phys.}\
  }\textbf {\bibinfo {volume} {4}},\ \bibinfo {pages} {144} (\bibinfo {year}
  {2008})}\BibitemShut {NoStop}%
\bibitem [{\citenamefont {Adam}\ \emph {et~al.}(2009)\citenamefont {Adam},
  \citenamefont {Brouwer},\ and\ \citenamefont {Das~Sarma}}]{Adam2009}%
  \BibitemOpen
  \bibfield  {author} {\bibinfo {author} {\bibfnamefont {S.}~\bibnamefont
  {Adam}}, \bibinfo {author} {\bibfnamefont {P.~W.}\ \bibnamefont {Brouwer}}, \
  and\ \bibinfo {author} {\bibfnamefont {S.}~\bibnamefont {Das~Sarma}},\ }\href
  {\doibase 10.1103/PhysRevB.79.201404} {\bibfield  {journal} {\bibinfo
  {journal} {Phys. Rev. B}\ }\textbf {\bibinfo {volume} {79}},\ \bibinfo
  {pages} {201404} (\bibinfo {year} {2009})}\BibitemShut {NoStop}%
\bibitem [{\citenamefont {Zhang}\ \emph {et~al.}(2009)\citenamefont {Zhang},
  \citenamefont {Brar}, \citenamefont {Girit}, \citenamefont {Zettl},\ and\
  \citenamefont {Crommie}}]{Zhang2009}%
  \BibitemOpen
  \bibfield  {author} {\bibinfo {author} {\bibfnamefont {Y.}~\bibnamefont
  {Zhang}}, \bibinfo {author} {\bibfnamefont {V.~W.}\ \bibnamefont {Brar}},
  \bibinfo {author} {\bibfnamefont {C.}~\bibnamefont {Girit}}, \bibinfo
  {author} {\bibfnamefont {A.}~\bibnamefont {Zettl}}, \ and\ \bibinfo {author}
  {\bibfnamefont {M.~F.}\ \bibnamefont {Crommie}},\ }\href {\doibase
  10.1038/nphys1365} {\bibfield  {journal} {\bibinfo  {journal} {Nat. Phys.}\
  }\textbf {\bibinfo {volume} {5}},\ \bibinfo {pages} {722} (\bibinfo {year}
  {2009})}\BibitemShut {NoStop}%
\bibitem [{\citenamefont {Van~Tuan}\ \emph {et~al.}(2016)\citenamefont
  {Van~Tuan}, \citenamefont {Ortmann}, \citenamefont {Cummings}, \citenamefont
  {Soriano},\ and\ \citenamefont {Roche}}]{vanTuan2016}%
  \BibitemOpen
  \bibfield  {author} {\bibinfo {author} {\bibfnamefont {D.}~\bibnamefont
  {Van~Tuan}}, \bibinfo {author} {\bibfnamefont {F.}~\bibnamefont {Ortmann}},
  \bibinfo {author} {\bibfnamefont {A.~W.}\ \bibnamefont {Cummings}}, \bibinfo
  {author} {\bibfnamefont {D.}~\bibnamefont {Soriano}}, \ and\ \bibinfo
  {author} {\bibfnamefont {S.}~\bibnamefont {Roche}},\ }\href
  {http://dx.doi.org/10.1038/srep21046} {\bibfield  {journal} {\bibinfo
  {journal} {Sci. Rep.}\ }\textbf {\bibinfo {volume} {6}},\ \bibinfo {pages}
  {21046} (\bibinfo {year} {2016})}\BibitemShut {NoStop}%
\bibitem [{\citenamefont {Kubo}(1957)}]{Kubo1957}%
  \BibitemOpen
  \bibfield  {author} {\bibinfo {author} {\bibfnamefont {R.}~\bibnamefont
  {Kubo}},\ }\href@noop {} {\bibfield  {journal} {\bibinfo  {journal} {J. Phys.
  Soc. Jpn.}\ }\textbf {\bibinfo {volume} {12}},\ \bibinfo {pages} {570}
  (\bibinfo {year} {1957})}\BibitemShut {NoStop}%
\bibitem [{\citenamefont {Fan}\ \emph {et~al.}(2014)\citenamefont {Fan},
  \citenamefont {Uppstu}, \citenamefont {Siro},\ and\ \citenamefont
  {Harju}}]{Fan2014}%
  \BibitemOpen
  \bibfield  {author} {\bibinfo {author} {\bibfnamefont {Z.}~\bibnamefont
  {Fan}}, \bibinfo {author} {\bibfnamefont {A.}~\bibnamefont {Uppstu}},
  \bibinfo {author} {\bibfnamefont {T.}~\bibnamefont {Siro}}, \ and\ \bibinfo
  {author} {\bibfnamefont {A.}~\bibnamefont {Harju}},\ }\href@noop {}
  {\bibfield  {journal} {\bibinfo  {journal} {Computer Physics Communications}\
  }\textbf {\bibinfo {volume} {185}},\ \bibinfo {pages} {28 } (\bibinfo {year}
  {2014})}\BibitemShut {NoStop}%
\bibitem [{\citenamefont {Settnes}\ \emph {et~al.}(2016)\citenamefont
  {Settnes}, \citenamefont {Leconte}, \citenamefont {Barrios-Vargas},
  \citenamefont {Jauho},\ and\ \citenamefont {Roche}}]{Settnes2016}%
  \BibitemOpen
  \bibfield  {author} {\bibinfo {author} {\bibfnamefont {M.}~\bibnamefont
  {Settnes}}, \bibinfo {author} {\bibfnamefont {N.}~\bibnamefont {Leconte}},
  \bibinfo {author} {\bibfnamefont {J.~E.}\ \bibnamefont {Barrios-Vargas}},
  \bibinfo {author} {\bibfnamefont {A.-P.}\ \bibnamefont {Jauho}}, \ and\
  \bibinfo {author} {\bibfnamefont {S.}~\bibnamefont {Roche}},\ }\href@noop {}
  {\bibfield  {journal} {\bibinfo  {journal} {2D Materials}\ }\textbf {\bibinfo
  {volume} {3}},\ \bibinfo {pages} {034005} (\bibinfo {year}
  {2016})}\BibitemShut {NoStop}%
\bibitem [{\citenamefont {Ferreira}\ and\ \citenamefont
  {Mucciolo}(2015)}]{Ferreira2015}%
  \BibitemOpen
  \bibfield  {author} {\bibinfo {author} {\bibfnamefont {A.}~\bibnamefont
  {Ferreira}}\ and\ \bibinfo {author} {\bibfnamefont {E.~R.}\ \bibnamefont
  {Mucciolo}},\ }\href@noop {} {\bibfield  {journal} {\bibinfo  {journal}
  {Phys. Rev. Lett.}\ }\textbf {\bibinfo {volume} {115}},\ \bibinfo {pages}
  {106601} (\bibinfo {year} {2015})}\BibitemShut {NoStop}%
\bibitem [{\citenamefont {Garc\'{\i}a}\ \emph {et~al.}(2015)\citenamefont
  {Garc\'{\i}a}, \citenamefont {Covaci},\ and\ \citenamefont
  {Rappoport}}]{Garcia2015}%
  \BibitemOpen
  \bibfield  {author} {\bibinfo {author} {\bibfnamefont {J.~H.}\ \bibnamefont
  {Garc\'{\i}a}}, \bibinfo {author} {\bibfnamefont {L.}~\bibnamefont {Covaci}},
  \ and\ \bibinfo {author} {\bibfnamefont {T.~G.}\ \bibnamefont {Rappoport}},\
  }\href {\doibase 10.1103/PhysRevLett.114.116602} {\bibfield  {journal}
  {\bibinfo  {journal} {Phys. Rev. Lett.}\ }\textbf {\bibinfo {volume} {114}},\
  \bibinfo {pages} {116602} (\bibinfo {year} {2015})}\BibitemShut {NoStop}%
\bibitem [{\citenamefont {Wei\ss{}e}\ \emph {et~al.}(2006)\citenamefont
  {Wei\ss{}e}, \citenamefont {Wellein}, \citenamefont {Alvermann},\ and\
  \citenamefont {Fehske}}]{Weisse2006}%
  \BibitemOpen
  \bibfield  {author} {\bibinfo {author} {\bibfnamefont {A.}~\bibnamefont
  {Wei\ss{}e}}, \bibinfo {author} {\bibfnamefont {G.}~\bibnamefont {Wellein}},
  \bibinfo {author} {\bibfnamefont {A.}~\bibnamefont {Alvermann}}, \ and\
  \bibinfo {author} {\bibfnamefont {H.}~\bibnamefont {Fehske}},\ }\href
  {\doibase 10.1103/RevModPhys.78.275} {\bibfield  {journal} {\bibinfo
  {journal} {Rev. Mod. Phys.}\ }\textbf {\bibinfo {volume} {78}},\ \bibinfo
  {pages} {275} (\bibinfo {year} {2006})}\BibitemShut {NoStop}%
\bibitem [{\citenamefont {Tal-Ezer}\ and\ \citenamefont
  {Kosloff}(1984)}]{TalEzer1984}%
  \BibitemOpen
  \bibfield  {author} {\bibinfo {author} {\bibfnamefont {H.}~\bibnamefont
  {Tal-Ezer}}\ and\ \bibinfo {author} {\bibfnamefont {R.}~\bibnamefont
  {Kosloff}},\ }\href@noop {} {\bibfield  {journal} {\bibinfo  {journal} {The
  Journal of Chemical Physics}\ }\textbf {\bibinfo {volume} {81}} (\bibinfo
  {year} {1984})}\BibitemShut {NoStop}%
\bibitem [{\citenamefont {Fehske}\ \emph {et~al.}(2009)\citenamefont {Fehske},
  \citenamefont {Schleede}, \citenamefont {Schubert}, \citenamefont {Wellein},
  \citenamefont {Filinov},\ and\ \citenamefont {Bishop}}]{Fehske2009}%
  \BibitemOpen
  \bibfield  {author} {\bibinfo {author} {\bibfnamefont {H.}~\bibnamefont
  {Fehske}}, \bibinfo {author} {\bibfnamefont {J.}~\bibnamefont {Schleede}},
  \bibinfo {author} {\bibfnamefont {G.}~\bibnamefont {Schubert}}, \bibinfo
  {author} {\bibfnamefont {G.}~\bibnamefont {Wellein}}, \bibinfo {author}
  {\bibfnamefont {V.~S.}\ \bibnamefont {Filinov}}, \ and\ \bibinfo {author}
  {\bibfnamefont {A.~R.}\ \bibnamefont {Bishop}},\ }\href {\doibase
  http://dx.doi.org/10.1016/j.physleta.2009.04.022} {\bibfield  {journal}
  {\bibinfo  {journal} {Phys. Lett. A}\ }\textbf {\bibinfo {volume} {373}},\
  \bibinfo {pages} {2182 } (\bibinfo {year} {2009})}\BibitemShut {NoStop}%
\bibitem [{\citenamefont {Xue}\ \emph {et~al.}(2011)\citenamefont {Xue},
  \citenamefont {Sanchez-Yamagishi}, \citenamefont {Bulmash}, \citenamefont
  {Jacquod}, \citenamefont {Deshpande}, \citenamefont {Watanabe}, \citenamefont
  {Taniguchi}, \citenamefont {Jarillo-Herrero},\ and\ \citenamefont
  {LeRoy}}]{Xue2011}%
  \BibitemOpen
  \bibfield  {author} {\bibinfo {author} {\bibfnamefont {J.}~\bibnamefont
  {Xue}}, \bibinfo {author} {\bibfnamefont {J.}~\bibnamefont
  {Sanchez-Yamagishi}}, \bibinfo {author} {\bibfnamefont {D.}~\bibnamefont
  {Bulmash}}, \bibinfo {author} {\bibfnamefont {P.}~\bibnamefont {Jacquod}},
  \bibinfo {author} {\bibfnamefont {A.}~\bibnamefont {Deshpande}}, \bibinfo
  {author} {\bibfnamefont {K.}~\bibnamefont {Watanabe}}, \bibinfo {author}
  {\bibfnamefont {T.}~\bibnamefont {Taniguchi}}, \bibinfo {author}
  {\bibfnamefont {P.}~\bibnamefont {Jarillo-Herrero}}, \ and\ \bibinfo {author}
  {\bibfnamefont {B.~J.}\ \bibnamefont {LeRoy}},\ }\href@noop {} {\bibfield
  {journal} {\bibinfo  {journal} {Nat Mater}\ }\textbf {\bibinfo {volume}
  {10}},\ \bibinfo {pages} {282} (\bibinfo {year} {2011})}\BibitemShut
  {NoStop}%
\bibitem [{\citenamefont {Ochoa}\ \emph {et~al.}(2012)\citenamefont {Ochoa},
  \citenamefont {Castro~Neto},\ and\ \citenamefont {Guinea}}]{Ochoa2012}%
  \BibitemOpen
  \bibfield  {author} {\bibinfo {author} {\bibfnamefont {H.}~\bibnamefont
  {Ochoa}}, \bibinfo {author} {\bibfnamefont {A.~H.}\ \bibnamefont
  {Castro~Neto}}, \ and\ \bibinfo {author} {\bibfnamefont {F.}~\bibnamefont
  {Guinea}},\ }\href {\doibase 10.1103/PhysRevLett.108.206808} {\bibfield
  {journal} {\bibinfo  {journal} {Phys. Rev. Lett.}\ }\textbf {\bibinfo
  {volume} {108}},\ \bibinfo {pages} {206808} (\bibinfo {year}
  {2012})}\BibitemShut {NoStop}%
\bibitem [{\citenamefont {Dyakonov}\ and\ \citenamefont
  {Perel}(1972)}]{Dyakonov1972}%
  \BibitemOpen
  \bibfield  {author} {\bibinfo {author} {\bibfnamefont {M.}~\bibnamefont
  {Dyakonov}}\ and\ \bibinfo {author} {\bibfnamefont {V.}~\bibnamefont
  {Perel}},\ }\href@noop {} {\bibfield  {journal} {\bibinfo  {journal} {Soviet
  Physics Solid State, Ussr}\ }\textbf {\bibinfo {volume} {13}},\ \bibinfo
  {pages} {3023} (\bibinfo {year} {1972})}\BibitemShut {NoStop}%
\bibitem [{\citenamefont {Boross}\ \emph {et~al.}(2013)\citenamefont {Boross},
  \citenamefont {D{\'o}ra}, \citenamefont {Kiss},\ and\ \citenamefont
  {Simon}}]{Boross2013}%
  \BibitemOpen
  \bibfield  {author} {\bibinfo {author} {\bibfnamefont {P.}~\bibnamefont
  {Boross}}, \bibinfo {author} {\bibfnamefont {B.}~\bibnamefont {D{\'o}ra}},
  \bibinfo {author} {\bibfnamefont {A.}~\bibnamefont {Kiss}}, \ and\ \bibinfo
  {author} {\bibfnamefont {F.}~\bibnamefont {Simon}},\ }\href@noop {}
  {\bibfield  {journal} {\bibinfo  {journal} {Scientific Reports}\ }\textbf
  {\bibinfo {volume} {3}},\ \bibinfo {pages} {3233} (\bibinfo {year}
  {2013})}\BibitemShut {NoStop}%
\bibitem [{\citenamefont {Konschuh}\ \emph {et~al.}(2010)\citenamefont
  {Konschuh}, \citenamefont {Gmitra},\ and\ \citenamefont
  {Fabian}}]{Konschuh2010}%
  \BibitemOpen
  \bibfield  {author} {\bibinfo {author} {\bibfnamefont {S.}~\bibnamefont
  {Konschuh}}, \bibinfo {author} {\bibfnamefont {M.}~\bibnamefont {Gmitra}}, \
  and\ \bibinfo {author} {\bibfnamefont {J.}~\bibnamefont {Fabian}},\
  }\href@noop {} {\bibfield  {journal} {\bibinfo  {journal} {Phys. Rev. B}\
  }\textbf {\bibinfo {volume} {82}},\ \bibinfo {pages} {245412} (\bibinfo
  {year} {2010})}\BibitemShut {NoStop}%
\bibitem [{\citenamefont {Abdelouahed}\ \emph {et~al.}(2010)\citenamefont
  {Abdelouahed}, \citenamefont {Ernst}, \citenamefont {Henk}, \citenamefont
  {Maznichenko},\ and\ \citenamefont {Mertig}}]{Abdelouahed2010}%
  \BibitemOpen
  \bibfield  {author} {\bibinfo {author} {\bibfnamefont {S.}~\bibnamefont
  {Abdelouahed}}, \bibinfo {author} {\bibfnamefont {A.}~\bibnamefont {Ernst}},
  \bibinfo {author} {\bibfnamefont {J.}~\bibnamefont {Henk}}, \bibinfo {author}
  {\bibfnamefont {I.~V.}\ \bibnamefont {Maznichenko}}, \ and\ \bibinfo {author}
  {\bibfnamefont {I.}~\bibnamefont {Mertig}},\ }\href@noop {} {\bibfield
  {journal} {\bibinfo  {journal} {Phys. Rev. B}\ }\textbf {\bibinfo {volume}
  {82}},\ \bibinfo {pages} {125424} (\bibinfo {year} {2010})}\BibitemShut
  {NoStop}%
\bibitem [{\citenamefont {Boettger}\ and\ \citenamefont
  {Trickey}(2007)}]{Boettger2007}%
  \BibitemOpen
  \bibfield  {author} {\bibinfo {author} {\bibfnamefont {J.~C.}\ \bibnamefont
  {Boettger}}\ and\ \bibinfo {author} {\bibfnamefont {S.~B.}\ \bibnamefont
  {Trickey}},\ }\href@noop {} {\bibfield  {journal} {\bibinfo  {journal} {Phys.
  Rev. B}\ }\textbf {\bibinfo {volume} {75}},\ \bibinfo {pages} {121402}
  (\bibinfo {year} {2007})}\BibitemShut {NoStop}%
\end{thebibliography}%

\end{document}